\pdfoutput=1
\documentclass[a4paper,11pt,twocolumn,nopdfoutputerror]{quantumarticle}
\newenvironment{ruledtabular}{\footnotesize\setlength{\tabcolsep}{4pt}\centering}{}
\providecommand{\keywords}[1]{}

\usepackage[utf8]{inputenc}
\usepackage[T1]{fontenc}
\usepackage{amsmath,amssymb,amsfonts}
\usepackage{graphicx}
\graphicspath{{figs_v2/}}
\usepackage{bm}
\usepackage{hyperref}
\usepackage{booktabs}
\usepackage{xcolor}

\newcommand{\avg}[1]{\langle #1 \rangle}
\newcommand{\Nc}{N_c}
\newcommand{\Nq}{N_q}
\newcommand{\Nshots}{N_{\mathrm{shots}}}
\newcommand{\MSE}{\mathrm{MSE}}
\newcommand{\QRA}{QRA}

\begin{document}

\title{Evaluation artifacts in reversible quantum reservoir protocols: a withdrawal of our own claims, and an entangled channel that survives the controls}

\author{Hikaru Wakaura}
\email{h.wakaura@qiri.co.jp}
\affiliation{QIRI (Quantum Integrated Research Institute Inc.), Tokyo 107-0061, Japan}

\author{Taiki Tanimae}
\email{t.tanimae@qiri.co.jp}
\affiliation{QIRI (Quantum Integrated Research Institute Inc.), Tokyo 107-0061, Japan}

\date{August 26, 2026}

\maketitle

\begin{abstract}
This replaces v1 of this manuscript, which reported that a quantum reservoir
autoencoder on reset-noise channels suppresses shot-noise sensitivity by ten
orders of magnitude and that a two-phase protocol reconstructs unseen inputs at
$\MSE \sim 10^{-4}$. \emph{We withdraw both
claims.} Under one consistent evaluation the suppression reverses sign: reset channels
degrade the reconstruction $144\times$, the apparent gain being an in-sample
number compared with an out-of-sample one. The
two-phase target is an element-wise map with a public key, invertible in closed
form without training at $\MSE = 7.4\times10^{-16}$, and a polynomial using no quantum
features matches or beats the full model in all 14 conditions.
We trace these to four evaluation artifacts; the fourth also affects the
companion work relied on here, whose latent code never leaves its
input-independent initialization, so substituting a constant changes nothing.

We then report a construction passing the same controls: an entangled channel
between two reservoir halves, injected on one side and read out on the other,
decoded by a readout calibrated once and applied to unseen messages. Over 32
realizations the held-out error is below chance in 31--32 (permutation
$p<10^{-4}$) while a shuffled control stays at chance; scrambler T-doping
proves necessary for one-shot injection, while interleaved injection
supplies its own non-Cliffordness. A classical scrambler of matched size
decodes the same task as well or better, so the transport is not evidence
of a quantum mechanism---the construction's claim is that the controls
pass, not that quantum helps. A decoder-free response-Gram spectrum exposes an access threshold invisible to
the decode error, though not a universal one: recomputed on nested
subsystems of one fixed circuit with right-censoring (repairing two defects
found in adversarial review), interleaved injection gives median
$f^{*}=0.25$--$0.33$, growing with the number of multiplexed components,
while one-shot injection reaches full detectability in only half or fewer
realizations at the largest settings. Linear visibility is necessary yet insufficient, diverging from decodability at
finite input amplitude---itself a design variable worth an order of magnitude. A variational ansatz trained to
convergence on the same task is beaten by the fixed untrained circuit by
$2.6$--$4.4\times$. Three operational extensions pass the same discipline:
arbitrary-length streaming requires dissipatively resetting the accessible
half after each readout (capacity $\approx0.9$ per chunk with per-position
decoding, $0.73$ under a single fixed decoder and a stationary key, versus
saturation near three chunks for unitary dynamics---at about twice the
per-chunk error of sending each chunk alone); the decoder does
not transfer across the eight scrambling keys tested (matched $0.14$,
mismatched $1.7$); and a
receiver holding the key matches the supervised ceiling to within a few
percent with no labelled transmissions, by calibrating on self-simulated
pairs. We close with the
controls, seed convergence and untrained baselines included, separating
transport from interpolation.
\end{abstract}

\section{Introduction and statement of withdrawal}\label{sec:intro}

Quantum reservoir computing couples fixed quantum dynamics to a trainable
linear readout~\cite{Fujii2017,Nakajima2021,Mujal2021,MartinezPena2021}, and its
appeal for near-term devices is that no gradients are propagated through a noisy
circuit~\cite{Cerezo2021,McClean2018,Holmes2022}. A recurring weakness is that
the classical controls needed to attribute performance to the \emph{quantum}
part of the pipeline are often absent~\cite{Schuld2019,Jaeger2001}.

The first version of this manuscript~\cite{v1} reported two results within a
quantum reservoir autoencoder (\QRA{})~\cite{Romero2017,Hinton2006}
instantiated with reset-noise channels~\cite{Dudas2023}:
a ten-order suppression of shot-noise sensitivity, and a two-phase protocol
that reconstructs previously unseen inputs. It also built on a companion
work~\cite{companion} that reported machine-precision reversible information
transformation.

We withdraw the two results of v1, and the companion work is being corrected
separately. The reason in each case is not a numerical error---every number in
v1 is reproducible from the same code---but an evaluation that could not
distinguish the claimed effect from an artifact of how the readout was fitted
and scored. We set out the four artifacts in
Secs.~\ref{sec:art1}--\ref{sec:art4}, and we state plainly that the primary
casualties are our own claims.

A negative result of this kind is only half of a contribution. The other half
is what we report in Secs.~\ref{sec:channel}--\ref{sec:wall}: having identified
\emph{why} the construction failed to transport information, we ask what would,
and test it under the same controls that killed the original. The answer is a
physical channel---an entangled joint reservoir, injected on one side and read
on the other---together with a decoder that is calibrated once rather than
refitted per message. It passes. We regard the pairing of the two halves, the
diagnosis and the repair judged by the same standard, as the point of this
paper.

\section{Artifact I: in-sample readout evaluation}\label{sec:art1}

\begin{figure}[htbp]
  \includegraphics[width=\columnwidth]{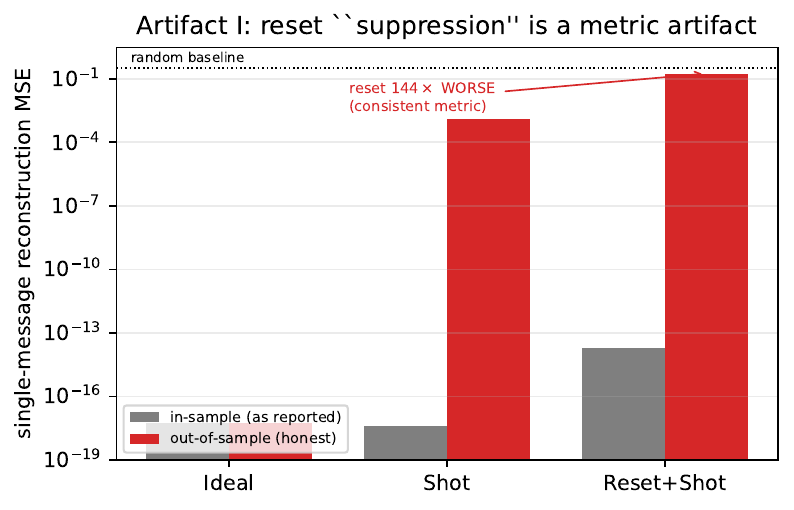}
  \caption{Artifact~I. In-sample (gray) reaches the machine floor for every
    noise model. Under a single consistent metric (red) the reset channels
    make the reconstruction $144\times$ \emph{worse}, approaching the random
    baseline. $\Nq=10$, $\Nc=10$.}
  \label{fig:art1}
\end{figure}

In the single-input mode of v1, decoder weights are obtained as
$W_{\mathrm{dec}} = \mathrm{solve}(V_{\mathrm{dec}}, \mathbf{C})$ and the
reconstruction is computed as
$\hat{\mathbf{C}} = V_{\mathrm{dec}} W_{\mathrm{dec}}$ using the \emph{same}
$V_{\mathrm{dec}}$. Since the feature dimension exceeds the target dimension,
the residual is an interpolation error at machine precision independently of
the noise model---the classic form of information leakage between fitting and
evaluation~\cite{Kapoor2023,Gundersen2018,Recht2019}. The honest metric re-measures: fit on $V_1$, evaluate on an
independent re-measurement $V_2$ of the same state
(Table~\ref{tab:oos}, Fig.~\ref{fig:art1}).

\begin{table}[htbp]
\caption{Single-input $\MSE$ at $\Nq=10$, $\Nc=10$ (median of 3--4 seeds).
  Under \emph{either} consistent metric, reset channels degrade the
  reconstruction. The reported ten-order suppression is the gap between the
  two columns, not an effect of the reset channels.}
\label{tab:oos}
\begin{ruledtabular}
\begin{tabular}{lcc}
Condition & In-sample & Out-of-sample \\
\hline
Ideal        & $5.4\times10^{-18}$ & $5.4\times10^{-18}$ \\
Shot         & $3.8\times10^{-18}$ & $1.23\times10^{-3}$ \\
Reset$+$shot & $2.0\times10^{-14}$ & $1.77\times10^{-1}$ \\
\hline
Effect of reset & $5300\times$ worse & $144\times$ worse \\
\end{tabular}
\end{ruledtabular}
\end{table}

The v1 shot value ($\sim\!10^{-3}$) matches our out-of-sample number and the
v1 reset$+$shot value ($\sim\!10^{-13}$) matches our in-sample number, so the
two were computed along different code paths. The proposed mechanism is also
internally inconsistent: reproducing an $10^{11}$-fold variance reduction via
$\sigma^2 = (1-\avg{Z}^2)/\Nshots$ requires $\avg{Z}$ within $10^{-11}$ of
unity, whereas the reset probabilities used are $p\sim\mathrm{U}(0,1)$ with
mean $0.48\pm0.28$---short of the claimed effect by roughly eight orders.

\section{Artifact II: the target is classically invertible}\label{sec:art2}

\begin{figure}[htbp]
  \includegraphics[width=\columnwidth]{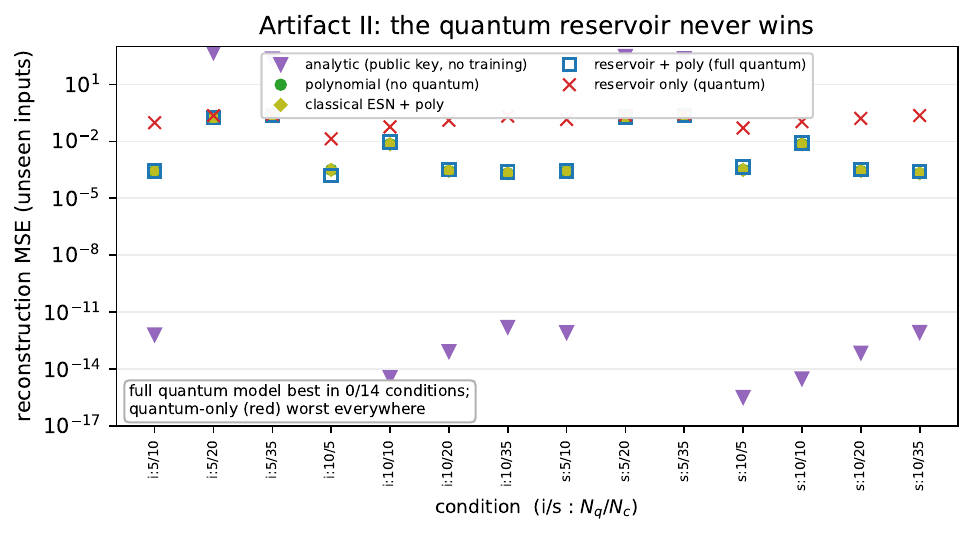}
  \caption{Artifact~II, all 14 conditions (labels i/s${:}\Nq/\Nc$). The full
    quantum model (open blue squares) coincides with the no-quantum polynomial
    and the classical ESN; the analytic public-key inverse (purple triangles)
    sits ten or more orders below every learned model wherever the task is
    solvable, showing the target is classically trivial.}
  \label{fig:art2}
\end{figure}

The reconstruction target of the two-phase protocol is the element-wise map
$\gamma_i = \tanh(A_i C_i + b_i)$, and (Sec.~\ref{sec:art3}) the latent code
equals it to $\sim\!10^{-8}$. Because $A$ is public, it inverts in closed form,
\begin{equation}
  C_i = \frac{\mathrm{atanh}(\gamma_i) - b_i}{A_i},
  \label{eq:analytic}
\end{equation}
with no training and no quantum device: on 100 unseen inputs at
$\Nq=\Nc=10$ this gives $\MSE = 7.4\times10^{-16}$, against
$2.2\times10^{-4}$ for the two-phase quantum protocol. Figure~\ref{fig:art2}
compares, on identical splits, the closed form, a per-position polynomial with
no quantum features, the reservoir features alone, the full reservoir$+$
polynomial model, and a classical echo state network~\cite{Jaeger2001,Maass2002,Lukosevicius2009}.
The full quantum model is best in \emph{none} of the 14 conditions, and the
quantum features alone are worst everywhere.

\section{Artifact III: structural noise immunity}\label{sec:art3}

The encoder is not a computation. For each message the encoding weights are
re-solved as
$W_{\mathrm{enc}} = \mathrm{solve}(V_{\mathrm{enc}}, F(A,\mathbf{C}))$, and
whenever the feature dimension is at least $\Nc$ this underdetermined system
reproduces its target: we measure
$\max_i |\gamma_i - F(A,\mathbf{C})_i| = 5.8\times10^{-9}$ at $\Nc = 10$.
The latent code is therefore insensitive to measurement noise \emph{by
construction}---whatever noise corrupts $V_{\mathrm{enc}}$, the re-solved
projection still lands on $F(A,\mathbf{C})$. This is why the analytic and
polynomial results are numerically identical between ideal and shot
conditions, and it is the origin of the ``no statistically significant
difference among noise conditions'' reported in v1. That invariance is a
property of re-solving a per-message projection, not evidence that a quantum
reservoir tolerates noise.

\section{Artifact IV: a latent code that carries nothing}\label{sec:art4}

The companion construction~\cite{companion} does not force
$\bm\gamma = F(A,\mathbf{C})$; it iterates $\bm\gamma$ from a random
initialization. Its own convergence analysis shows the iterate update is a
projection $P_a$ with unit eigenvalues on the row space, so
$P_a\bm\gamma^{(0)} = \bm\gamma^{(0)}$---described there as ``essentially a
no-op.'' The consequence, not drawn there, is that $\bm\gamma$ never leaves an
initialization drawn independently of the input.

\begin{figure}[htbp]
  \includegraphics[width=\columnwidth]{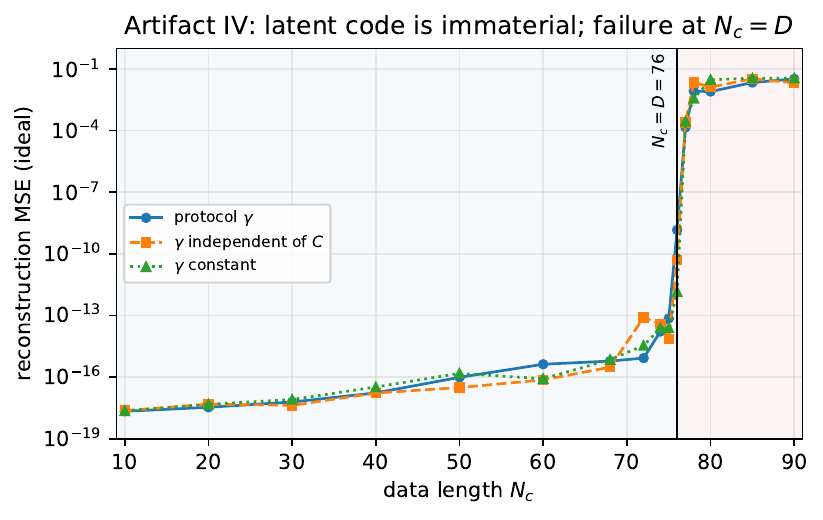}
  \caption{Artifact~IV, companion construction ($d=76$, ideal). The
    reconstruction is indistinguishable whether the latent code is the
    protocol's own, drawn independently of the input, or constant---the three
    curves lie on top of one another---and all jump by fourteen orders at the
    encoder rank threshold $\Nc = d = 76$.}
  \label{fig:art4}
\end{figure}

Using that implementation unmodified, replacing the latent code by a vector
drawn independently of $\mathbf{C}$, or by a constant, leaves the
reconstruction unchanged at the double-precision floor
(Fig.~\ref{fig:art4}); the iterate does not move
($|\bm\gamma^{(1)}-\bm\gamma^{(0)}|_{\max}\approx10^{-9}$), and a held-out
decoder trained on $(\bm\gamma,\mathbf{C})$ pairs recovers the input no better
than chance ($0.272$ against $\mathrm{Var}(\mathbf{C}) = 0.271$ at $\Nc=10$;
the same test on a genuine element-wise codec gives $2.5\times10^{-3}$, so the
test does detect a channel when one exists). The machine precision is
single-message memorization by an underdetermined ridge
readout~\cite{Tikhonov1963,Hoerl1970}, and it vanishes at
the encoder rank threshold $\Nc = d$.

\section{A channel that passes the controls}\label{sec:channel}

The four artifacts have a common root, and separating it into two independent
failures is what makes a repair possible. The first is that the latent code
does not depend on the input. The second is that the decoder is handed the
reconstruction target during training. Fixing either alone is useless: a
genuine channel whose decoder is still given $\mathbf{C}$ reproduces the same
uninformative machine precision, and an honest decoder fed an inert code
simply fails.

We therefore replace the classical latent vector by a physical channel. Two
reservoir halves are evolved as one joint scrambling system of $n$ qubits: the
data are injected on the inaccessible side (one component per qubit, so that
components cannot compose into a single rotation), the accessible side
$A$ of $a = fn$ qubits is read out through its low-weight Pauli expectations,
and the decoder is fit once on training messages and then applied to messages
it has never seen. Restricting the readout to weight $\leq 2$ keeps the
feature dimension ($3a + 9\binom{a}{2}$; $105$ at $a=5$) below the final
calibration set of $120$ messages, though not below the $80$-message fit
split used for selecting the ridge parameter---at the main operating point
the regularization is therefore chosen in a mildly underdetermined regime,
with the held-out block untouched throughout. An earlier version of this
sentence claimed the dimension stayed below the training count
``throughout''; adversarial review found the counts above, and we state
them instead.

\begin{figure}[htbp]
  \includegraphics[width=\columnwidth]{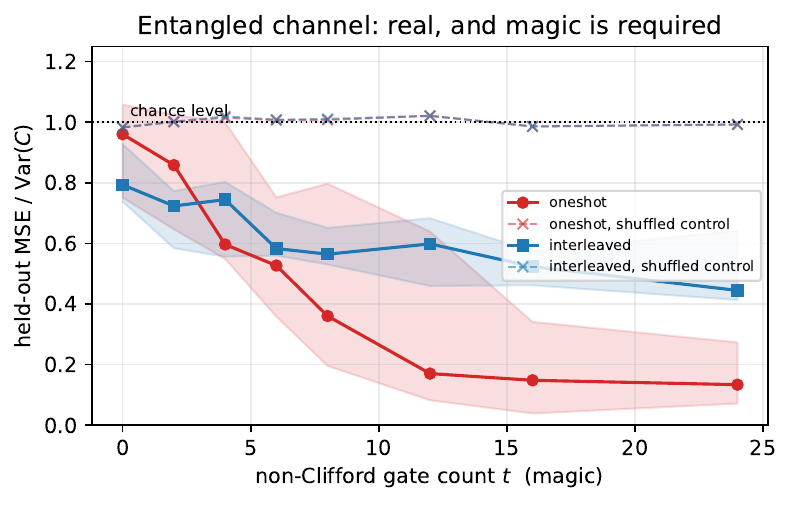}
  \caption{The entangled channel transports the input. Held-out reconstruction
    error, normalized by $\mathrm{Var}(C)$, versus non-Clifford gate count;
    bands span six circuit realizations. Crosses are the shuffled control,
    which stays at chance throughout, so the decoder is not exploiting
    anything but the channel. At $t=0$ (Clifford scrambler) the result at
    this amplitude is chance; the amplitude-resolved scan in the text shows
    the necessity of scrambler doping is specific to one-shot injection,
    not a blanket property.}
  \label{fig:channel}
\end{figure}

Figure~\ref{fig:channel} reports the held-out reconstruction against the
number $t$ of non-Clifford gates, at $n=10$, $f=1/2$, $\Nc=4$, over six random
circuit realizations, with a shuffled negative control in which the
(input, features) pairing is destroyed in the training split only.

Two things follow. The channel is real---the held-out error falls below
chance for both injection schemes while the shuffled control stays at
$0.98$--$1.02$ everywhere. (The figure shows six realizations at a single
amplitude; Sec.~\ref{sec:baseline} reports the same measurement over 32
realizations, where the median, $0.51$, is higher than these six suggest,
together with the amplitude dependence that dominates it.) And the role of
non-Clifford resources is sharper than we first stated---in both
directions. Adversarial review challenged the blanket claim ``magic is
necessary'', so we scanned doping against amplitude and injection mode,
with linear and quadratic readouts, over eight realizations
(\texttt{magic\_scan.json}). For one-shot injection the claim survives
every probe: at $t=0$ the held-out decoder sits at chance ($0.990$) at
every amplitude from $0.1$ to $0.9$, quadratic features included, and the
decoder-free response rank is zero in most realizations---without scrambler
doping the input leaves no detectable low-weight trace on $A$ at all. For
interleaved injection it does not survive: at $t=0$ and amplitude $0.3$ the
held-out decoder reaches $0.49$. Half the variance crosses a purely
Clifford scrambler, because the repeatedly injected rotations
$e^{-i C_i X}$ are themselves non-Clifford gates and supply what the
scrambler lacks. The correct statement is therefore mode-specific:
T-doping of the \emph{scrambler} is necessary when the data enter once;
when the data re-enter at every block, the injection provides its own
non-Cliffordness and doping ``only'' buys another two orders of magnitude
($t=12$, amplitude $0.3$: $0.004$). Scrambling alone still opens no
channel anywhere. This is consistent with the role of
non-stabilizerness as the resource separating classically simulable scrambling
from generic dynamics~\cite{Leone2022,Haferkamp2022}, with the decoder
fidelity for quasi-chaotic scramblers being governed by the same
resource~\cite{Leone2024}, and with our own companion study of identifiability
under doped-Clifford scrambling~\cite{qnoise}.

\paragraph{What the name does and does not claim.}
Adversarial review asked whether ``entangled channel'' names a mechanism or
a state. Measured: the $A|B$ entanglement entropy of the encoded state is
near-maximal (median $4.3$ of $5$) at every doping level including $t=0$,
and across 32 realizations it is uncorrelated with the held-out error
(Pearson $-0.03$ one-shot, $-0.08$ interleaved). Entanglement is generic
here and is not the discriminating resource---the circuits that fail are as
entangled as the circuits that work. The converse control is also
measured: a bare qubit-routing circuit---a permutation moving the
injection qubits into $A$, no other gates---decodes at $0.038$ with
\emph{exactly zero} $A|B$ entanglement (\texttt{swap\_control.json}).
Transport needs no entanglement at all. The name therefore describes the
state the information crosses, not a mechanism the controls have
isolated: the channel transports \emph{through} a fully entangling
dynamics, not \emph{because of} one, and the resource that separates
working from failing scramblers is the T-doping.

\paragraph{The classical control, finally applied to ourselves.}
Our Introduction faults the field for attributing performance to the
quantum part of a pipeline without a classical control, and Artifact~II
kills v1 with exactly such a control. Adversarial review observed that the
repaired channel had none---the shuffled control tests whether \emph{any}
signal exists, not whether a quantum mechanism carries it. Added here: the
same task (inject $\Nc$ components on a hidden half, read the other half
through units and their pairwise products, same calibrated ridge, same
held-out protocol, same shot noise) on a classical scrambler of $2n$ units
(four layers of random orthogonal mixing with elementwise $\tanh$) and on a
leaky echo-state reservoir~\cite{Jaeger2001,Maass2002,Lukosevicius2009}. Over 32 realizations the classical medians are
$0.007$/$0.008$ (orthogonal-$\tanh$, one-shot/re-injected) and
$0.202$/$0.034$ (ESN)---as good as or far better than the quantum channel's
$0.513$/$0.616$ at the same amplitude, and better than its
amplitude-optimized $\approx 0.08$. The conclusion our own standard forces:
\emph{the transport demonstrated here is not evidence of a quantum
mechanism.} The construction shows that the encode--decode geometry can
carry information while every artifact control passes---which is what v1
failed to show---not that a quantum system does it better than a classical
one. The Discussion already declined to claim an advantage; the classical
control converts that disclaimer into a measurement. The gap itself is a
measurement-budget effect, not a ceiling, and we quantify it under a fully
physical accounting: the total number of state preparations $N_{\rm tot}$
is split across qubit-wise-commuting measurement settings (a greedy cover
packs the 105 weight-$\leq2$ Paulis at $a=5$ into 16 settings), each Pauli
carrying state-dependent binomial variance $(1-\langle P\rangle^2)/N_s$;
the classical control reads all of its features in a single
``setting''---everything commutes---which is precisely its structural
advantage. At the channel's best operating point (interleaved, amplitude
$0.2$, $t=12$) the held-out ratio falls
$0.375/0.066/0.012/0.0057/0.0030$ at
$N_{\rm tot} = 10^4/10^5/10^6/3\times10^6/10^7$, while the classical
control saturates at its own floor of $0.0039$ from
$N_{\rm tot}=3\times10^5$ on: \emph{the quantum channel matches the
classical control near $N_{\rm tot} \approx 5\times10^6$ preparations and
passes it at $10^7$}, with grouping itself worth $4$--$6\times$ over
splitting the budget per observable
(\texttt{shot\_grouping.json}; per-observable levers---amplitude, mode,
$M$, best-of-eight key selection---are swept in
\texttt{quantum\_accuracy.json}). Scrambling and non-commutativity dilute
each preparation's information, so the channel pays for its access
threshold and key sensitivity in preparations; the classical control,
which hides nothing and commutes everywhere, pays nothing---but it also
has a floor, and the channel does not (exact-expectation limit $0.0001$).

\paragraph{Hardware.}
The channel runs on hardware. On \texttt{ibm\_fez} (Heron, 156 qubits),
at $n=6$, $\Nc=2$, $a=3$, one-shot injection, amplitude $0.3$, $M=40$
messages, all 36 weight-$\leq2$ Paulis in 9 qubit-wise-commuting settings
(720 circuits, $10^3$ shots each, 194\,s of QPU time, median two-qubit
depth 100 after transpilation), the decoder calibrated on the device's own
(message, feature) pairs reaches a held-out ratio of $0.23$, against
$0.035$ for the identical circuits in noiseless simulation, with the
shuffled control at chance ($1.19$): device noise costs a factor of
$\sim6.5$ and is absorbed into calibration rather than fatal. The $t=0$
control is the surprise. At chance in simulation ($1.18$), it decodes at
$0.67$ on the device---coherent gate errors are themselves uncontrolled
non-Clifford resources, partially opening the channel that exact Clifford
dynamics keeps closed. That is a hardware caveat for anything that relies
on the $t=0$ closure, and an unplanned confirmation of the doping scan's
lesson that small non-Clifford perturbations suffice
(\texttt{hardware\_channel.json}).

\section{Seed statistics and a trained-circuit baseline}\label{sec:baseline}

Two questions decide whether Sec.~\ref{sec:channel} is a result or an anecdote:
whether it survives seed statistics, and whether a circuit trained for the task
does better.

\paragraph{Seed statistics.}
Repeating the channel measurement over 32 circuit realizations
(Table~\ref{tab:repro}), the held-out error is below chance in $31/32$
realizations for one-shot injection and $32/32$ for interleaved, with a
two-sided permutation test against the shuffled control giving $p < 10^{-4}$ in
both cases. The medians, however, drift with the number of realizations: for
one-shot the median over the first $4$, $8$, $16$ and $32$ seeds is
$0.346$, $0.346$, $0.422$, $0.513$. An eight-seed estimate looked stable and
was not---it was optimistic by a factor of $1.5$. We therefore report 32-seed
medians with inter-quartile ranges throughout, and we note that our own first
pass at this measurement would have overstated the channel.

\begin{table}[htbp]
\caption{Channel measurement over 32 realizations at $n=10$, $f=1/2$, $t=12$,
  amplitude $0.9$. The shuffled control destroys the (input, feature) pairing
  in the training split only.}
\label{tab:repro}
\begin{ruledtabular}
\begin{tabular}{lccc}
 & median [IQR] & shuffled & below chance \\
\hline
one-shot    & $0.513$ $[0.34, 0.67]$ & $1.009$ & $31/32$ \\
interleaved & $0.616$ $[0.56, 0.68]$ & $1.016$ & $32/32$ \\
\end{tabular}
\end{ruledtabular}
\end{table}

\paragraph{Amplitude is a first-class design variable.}
The amplitude sweep of Sec.~\ref{sec:wall} is not a marginal effect. At the
optimum the same configuration that gives $0.62$ at amplitude $0.9$ gives
$0.08$, and the per-realization optimum is interior to the grid in $97$--$100\%$
of seeds, concentrated at $0.2$. A protocol reported at a single amplitude can
understate itself by an order of magnitude.

\paragraph{A trained circuit does worse.}
The natural competitor is a circuit optimized for the task rather than fixed:
a hardware-efficient RY/ring-CZ ansatz of the kind used for quantum
autoencoders~\cite{Romero2017}, with $30$ parameters, trained by COBYLA on the
training split only. To separate learning from the ansatz family we also score
the same ansatz at its random initialization---without that control, an
advantage of the family would be indistinguishable from an advantage of
training.

\begin{table}[htbp]
\caption{Same task, same readout, same calibrated decoder, and---after an
  adversarial-review repair---the same run: all three circuits are scored
  on one message set per seed, $M=120$, 16 seeds, $n=10$, $f=1/2$,
  amplitude $0.3$ (an earlier version of this table mixed two runs with
  different $M$ and seed counts, exactly the provenance error this paper
  documents; \texttt{baseline\_unified.json}). The trained ansatz is
  scored at 150/300/600 objective evaluations; its median is flat
  ($0.288/0.288/0.284$ one-shot), which is the convergence evidence.}
\label{tab:qae}
\begin{ruledtabular}
\begin{tabular}{lccc}
Circuit & one-shot & interleaved & circuit evals \\
\hline
$t$-doped Clifford, fixed  & $\mathbf{0.108}$ & $\mathbf{0.068}$ & $0$ \\
variational, untrained     & $0.559$ & $0.558$ & $0$ \\
variational, trained       & $0.284$ & $0.296$ & $600$ \\
\end{tabular}
\end{ruledtabular}
\end{table}

Training works---it improves the ansatz by roughly a factor of two over its
initialization---and it is still beaten by the untrained fixed circuit, by
$2.6\times$ for one-shot and $4.4\times$ for interleaved
(Table~\ref{tab:qae}). The optimizer has converged in the only sense we can
defend: the held-out median is unchanged across 150, 300 and 600 objective
evaluations, in the same run that produced the table. Six hundred circuit
evaluations correspond to tens of
thousands of circuit executions on hardware, spent to lose to a circuit that
required none.

We are deliberate about what this does and does not say. It is a statement
about one ansatz family, one gradient-free optimizer, one initialization per
seed, and one system size; a deeper ansatz, a gradient-based method, or
restarts could change it, and the early flattening under interleaved injection
is consistent with a local optimum rather than a representational limit. What
it does establish is that the fixed-circuit result is not a weak baseline
dressed up: the obvious trained alternative, run to convergence, does worse.

\section{An access threshold, and two ways to be wrong about
it}\label{sec:wall}

Whether information reaches $A$ at all is an information-theoretic question,
and asking it through a decoder confounds the channel with the readout's
capacity. We therefore also measure a decoder-free quantity. Perturbing each
input component,
$R_i = [\,\mathrm{feat}(+\delta e_i) - \mathrm{feat}(-\delta e_i)\,]/2\delta$,
the response Gram $\Gamma_{ij} = \langle R_i, R_j\rangle$ has full rank if and
only if every direction of the input leaves an independent first-order trace
in the measured observables; this is the classical-Fisher analogue of the
quantum Fisher information~\cite{Braunstein1994} restricted to $A$ and to
the weight-$\leq2$ Pauli set. Nothing is fitted. To make the count
operational rather than formal we score each response direction against a
stated per-observable budget, calling it detectable when
$\lambda_k \mathrm{Var}(C) N > 1$ at $N$ shots per expectation value; this
is an idealization---it charges every Pauli the same $1/N$ variance and
does not allocate a total copy budget across measurement settings, a
convention that flatters large $a$ in principle; the Limitations report
the physical-budget recomputation, which leaves the thresholds
unchanged. A purely relative rank
tolerance instead counts directions whose absolute response is $10^{-27}$,
which we found reports full visibility at cells where the decoder sits at
chance.

\begin{figure*}[tbp]
  \includegraphics[width=\textwidth]{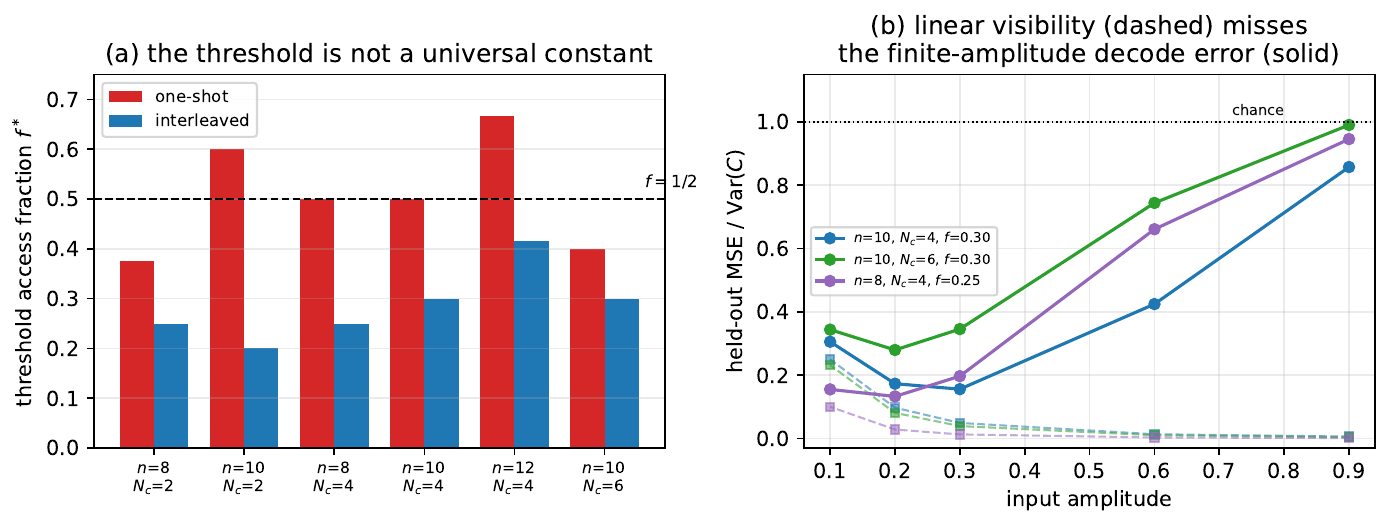}
  \caption{(a)~The threshold access fraction $f^{*}$ at which all $\Nc$
    components become detectable ($N = 10^4$ shots, $t = 12$, eight circuit
    realizations). This panel shows the \emph{original} scan, in which each
    access fraction used an independently drawn circuit; the
    nested-subsystem recomputation with right-censoring (text, and
    \texttt{threshold\_nested.json}) confirms the qualitative
    pattern---$f^{*}$ grows with $\Nc$ and interleaved injection lowers
    it---while revising the one-shot medians upward and censoring the
    largest settings entirely. (b)~Held-out decode error (solid) against the
    prediction from linear response (dashed), versus input amplitude. The two
    agree only as the amplitude vanishes; at amplitude $0.9$ they differ by up
    to $0.95$. The decode error is non-monotonic, with an optimum near
    $0.2$--$0.3$.}
  \label{fig:threshold}
\end{figure*}

\paragraph{The threshold is real where it exists---and it does not exist
  everywhere.}
Adversarial review found two defects in our original threshold analysis;
both are repaired here, and one conclusion does not survive the repair.
First, the original scan seeded a fresh scrambler at every access fraction,
so a ``per-realization threshold'' was read off a staircase of unrelated
random circuits---and that staircase violates the monotonicity that access
scanning requires (the detectable count must be non-decreasing when more of
the \emph{same} state is read; the old scheme failed this check in most
realizations). Recomputed on nested subsystems
$A_1 \subset A_2 \subset \cdots$ of one fixed circuit per realization, where
the response Gram grows in PSD order and monotonicity is guaranteed, the
check passes in $32/32$ realizations everywhere. Second, realizations that
never reach full rank within the accessible range are right-censored, not
missing; excluding them (as our earlier aggregation did) biases $f^{*}$
low. With both repairs, at $t=12$ and $10^{4}$ shots per expectation:
interleaved injection has a genuine threshold, median $f^{*}=0.25$--$0.33$
across $n=8$--$12$, $\Nc=2$--$6$, with at most $4/32$ realizations
censored. One-shot injection has one only at the smaller settings (median
$0.31$--$0.50$); at $(n,\Nc)=(10,6)$ and $(12,4)$ only $11/32$ and $16/32$
realizations reach full rank at \emph{any} accessible fraction, so the
honest summary there is $f^{*}>0.40$ and $f^{*}>0.67$---for most such
circuits some input direction stays undetectable no matter how much of the
system is read. Our earlier statement that a threshold exists in every
setting is withdrawn accordingly. Below the threshold (where one exists)
some direction of the input leaves no detectable first-order trace at the
stated budget, so no estimator reading these observables at this budget
recovers it at first order; higher-order response is not probed, so
``impossible in principle'' would overstate it. The position, where
defined, depends on the setting
[Fig.~\ref{fig:threshold}(a)], increasing with $\Nc$; interleaved
injection gives a lower threshold at every setting tested, and at
$n=10$, $\Nc=4$ the inter-quartile ranges of the two do not overlap. The
growth of $f^{*}$ with $\Nc$ is the direction a Page/Hayden--Preskill
argument suggests, and the quantitative comparison separates the two: the
decoupling count for recovering $k$ qubits of \emph{quantum} information
from a random unitary on $n$ qubits requires reading just over $(n+k)/2$
of them, $f \gtrsim 1/2 + k/2n$~\cite{Hayden2007}---for
$(n,\Nc)=(8,2)$, $(8,4)$, $(10,4)$ that is $0.63$, $0.75$, $0.70$,
while the measured one-shot medians are $0.31$, $0.38$, $0.50$. The
observed thresholds sit at roughly half the Hayden--Preskill requirement,
as they should: the task asks for $\Nc$ classical parameters at first
order, not the quantum state. $f^{*}(\Nc)$ is a classical-parameter
detectability threshold, distinct from---and much cheaper than---HP
recovery. In an earlier
version of this analysis we examined the single case $n = 10$, $\Nc = 4$,
found $f^{*} = 1/2$, and described it as a sharp no-cloning wall. That
identification was too strong: a no-cloning bound on a scalar signal should not
move with the number of multiplexed components, and ours does. What we measure
is the access needed to resolve $\Nc$ components simultaneously, which
coincides with $1/2$ only incidentally. Interleaved re-injection consistently
lowers $f^{*}$ for $\Nc \geq 4$, in agreement with the qualitative
expectation that a persistent signal is easier to recover than a one-shot one.

\paragraph{Linear visibility does not imply decodability.}
The decoder-free metric is necessary but not sufficient, and the gap is not
small. Figure~\ref{fig:threshold}(b) shows cells where every component is
comfortably detectable in linear response---the prediction
$\Nc^{-1}\sum_k (1+\mathrm{SNR}_k)^{-1}$ sits below $0.01$---while the
measured held-out error is at chance. Over all 72 cells the mean discrepancy
is $0.31$ for a rank-counting predictor and $0.34$ for the continuous one,
with correlations of $+0.66$ and $+0.67$: enough to be related, nowhere near
enough to be a law. The cause is amplitude. Linear response is a
small-signal quantity, whereas the protocol injects components of order unity,
and under interleaved encoding those injections compose through repeated
rotations rather than adding. Reducing the amplitude collapses the discrepancy
monotonically, from $0.85$ at amplitude $0.9$ to $0.06$ at $0.1$, confirming
the mechanism.

\paragraph{An optimal input amplitude.}
The same figure shows a design consequence we did not anticipate. The held-out
error is non-monotonic in amplitude: large amplitudes are destroyed by
nonlinear composition, small ones sink below the shot-noise floor, and the
optimum lies near $0.2$--$0.3$, where the error falls from chance to
$0.13$--$0.28$. Input amplitude is therefore a design variable of the same
standing as access fraction and magic, and reporting a protocol's performance
at a single amplitude---as we did in v1---can understate it by a factor of
several.

\paragraph{What this costs the earlier claim.}
Two of the three statements we would have made from the single-setting scan do
not survive: the threshold is not a universal $1/2$, and the decoder-free
metric is not a proxy for decodability. The one that does survive is the
methodological one, and it is the reason we report both quantities: the decode
error alone shows no feature at the threshold, so a study reporting only
reconstruction accuracy---the natural analogue of what v1 reported---would
conclude that the channel degrades gracefully where in fact some component of
the input never arrives at all.

\section{Operational extensions: streaming, keys, and label-free
  calibration}\label{sec:extensions}

A channel that passes the controls invites three questions it was not built to
answer: whether it extends beyond fixed-length messages, whether it is keyed,
and whether the calibration can avoid labelled transmissions altogether. We
measured all three, at $n=8$, $f=1/2$, with the same discipline as above---every
number is a held-out error normalized by $\mathrm{Var}(C)$, medians over random
circuit realizations, with the appropriate negative control run alongside.

\paragraph{Streaming requires dissipation.}
A length-$L$ sequence of two-component chunks is injected one chunk per slot,
each followed by a full segment of scrambling (four blocks, $t=16$). Reading
$A$ once at the end recovers about two chunks and collapses entirely by $L=8$
(Table~\ref{tab:streaming}). Reading $A$ after every segment while keeping the
dynamics unitary does no better asymptotically: the first two or three chunks
decode and the capacity then plateaus near three chunks regardless of $L$. The
reservoir is unitary, so nothing it absorbs ever leaves; the accumulated
history acts as a message-dependent background, and---unlike a classical
echo-state network---there is no fading memory to age it out. That a
quantum reservoir needs non-unitarity for fading memory is established in
the quantum-reservoir-computing
literature~\cite{Fujii2017,Nakajima2021,Mujal2021}; the contribution of
Table~\ref{tab:streaming} is not that principle but its quantitative form
for this channel---the unitary saturation capacity ($\approx 3$ chunks,
independent of $L$) and the per-chunk capacity the reset recovers
($\approx 0.9$). Giving the
decoder the previous readout as covariates does not remove the background
(the low-weight Pauli expectations on $A$ do not determine it), which locates
the obstruction beyond the reach of that decoder class (nonlinear readouts
were not tried). The cure is
dissipative: resetting the accessible half to $|0\rangle$ after each readout
(simulated as measure-and-discard, in density-matrix evolution) keeps the
per-chunk error bounded---non-growing in position and $L$, though scattered
across positions by segment-to-segment variance ($0.03$--$0.26$ at sixteen
realizations)---with
capacity $\approx 0.9$ chunks per chunk sent out to $L=16$. Two disclosures
matter for reading Table~\ref{tab:streaming}: each position is decoded by
its own ridge head, and the segments are drawn independently at every step,
a time-varying key schedule---so the table by itself demonstrates arbitrary
length only for pre-calibrated positions. The stationary-key variant closes
that gap. With the same segment applied at every step the per-position
capacity is if anything better ($15.1$ at $L=16$, eight realizations), and
a \emph{single} head, calibrated at one position of an $L=8$ run and
applied frozen to every position past the two-step transient, holds at
$0.27$ per chunk with fresh messages at $L=16$ \emph{and} at
$L=32$---$0.272$ versus $0.274$, no drift with length, $0.73$ of each
chunk's variance recovered. (The two transient positions decode far worse
than chance, $49$ and $3.9$; a deployment discards them.) Arbitrary length
under one fixed decoder therefore stands for a stationary key out to twice
the calibrated length, at roughly twice the per-chunk error of
position-specific heads.

\begin{table}[htbp]
\caption{Streaming capacity, $\sum_k \max(0,\,1-r_k)$ over chunk positions
  $k$ with $r_k$ the held-out ratio (positions decoding worse than chance
  contribute zero), versus sequence length $L$, at $n=8$, $f=1/2$, two
  components per chunk, four-block $t=16$ segments, $M=200$, medians over
  sixteen realizations. Each position is decoded by its own ridge head,
  calibrated once on the training split. A unitary reservoir saturates;
  resetting $A$ after each readout scales.}
\label{tab:streaming}
\begin{ruledtabular}
\begin{tabular}{lcccc}
 & $L=2$ & $L=4$ & $L=8$ & $L=16$ \\
\hline
single final readout & 1.87 & 1.18 & 0.01 & 0.05 \\
streaming, unitary & 1.91 & 2.71 & 2.73 & 2.85 \\
streaming, $A$ reset & 1.86 & 3.39 & 6.94 & 14.42 \\
\end{tabular}
\end{ruledtabular}
\end{table}

Neither inserted operation can be thinned. Ablating at $L=8$ (steady-state
ratio over positions $2$--$7$, sixteen realizations; full protocol
$0.17$): removing the reset gives
$0.95$; halving the segment doping to $t=8$ gives $0.94$ and $t\le4$ is
chance, consistent with the doping requirement of
Sec.~\ref{sec:channel}; halving the segment depth to two blocks gives $0.69$
and one block is chance. At sixteen realizations the separation between
the full protocol and every ablation is clean, though the $t=8$ cell still
spans $0.53$--$1.08$ across realizations, so the threshold's exact
location in $t$ remains realization-dependent. Decoder-side improvements
then stack: a two-step
readout window with squared features and doubled calibration data reaches
$0.120$, and letting the legitimate parties select the best of four
candidate keys on a validation batch---scored on a fresh message
batch---reaches $0.066$ (median key $0.153$), with no further improvement
resolved when stacking continues (three realizations for the stacked
cells). The honest baseline for that number is the payload-matched
static channel: a single-shot transmission of one two-component chunk
decodes at $0.031$ (median over 32 keys; best-of-four selection gains
nothing there, $0.031$, because the key-to-key variance at $\Nc=2$ is
already small---the selection gain is a property of high-variance settings,
not of streaming). Streaming thus costs about a
factor of two per chunk relative to sending each chunk alone---the residual
price of the recycled reservoir---in exchange for amortizing one circuit
over the whole sequence.

\paragraph{The channel is keyed.}
Taking the key to be the seed that synthesizes the doped-Clifford blocks, a
decoder calibrated under key $i$ and applied to transmissions under key $j$
gives, over an $8\times8$ key matrix and eight realizations ($t=16$,
one-shot, $\Nc=4$), a matched-key median of $0.140$ against a mismatched
median of $1.72$, with every one of the 56 off-diagonal cells above
$1.3$---worse than chance, a wrong-key decoder adds noise rather than
extracting signal. Within the scope measured---eight keys, a single-key
ridge decoder---calibration does not transfer; multi-key joint decoders,
key estimation, and chosen-plaintext attacks are untested, so this is a
transfer statement, not a keyspace analysis (reservoir-based constructions
have been proposed in post-quantum-cryptography
contexts~\cite{Cossins2024}; the statement here is far narrower than a
security claim). The
honest caveat is measured alongside: an eavesdropper with access to all $n$
qubits \emph{and} the labelled calibration pairs decodes at $0.039$, so the
key protects only while the pairs are secret or the access fraction is below
the wall of Sec.~\ref{sec:wall}. The next paragraph removes the pair-supply
route for this caveat, though not the access condition.

\paragraph{Label-free calibration matches the supervised ceiling to within
  a few percent.}
The decoder above needs $M$ labelled (message, feature) pairs. A receiver who
holds the key needs none of them from the sender: it can synthesize the
blocks, simulate its own labelled pairs, and calibrate on those---a digital
twin. The comparison is deliberately symmetric: twin and supervised decoder
use an identical calibration budget ($40\%$ fit, $20\%$ validation, refit
on both) and are evaluated on the identical held-out block of the
transmission batch, so the two numbers differ only in where the labels came
from. On that footing ($M=800$, four realizations), the twin-calibrated
decoder gives $0.142$ against the supervised $0.133$ (interleaved: $0.518$
against $0.515$; with four realizations no difference is resolved, though
formal equivalence is not established), and degrades to $0.218$
when the received features carry $10^4$-shot noise while the calibration
stays noiseless. A $t=0$ control
stays at chance ($0.95$), so the twin is decoding the channel, not a
simulation artifact. The word ``twin'' is quantified rather than assumed:
when the transmitter carries coherent $Z$-errors of $\varepsilon$ rad per
qubit per block that the twin's model does not know about, the ratio
degrades as $0.152 \to 0.159/0.209/0.414/1.27$ at
$\varepsilon = 0.01/0.03/0.1/0.3$ (four realizations)---the receiver's
model must be accurate at the few-$\times 10^{-2}$-rad level, and beyond
$\varepsilon \approx 0.1$ simulator-generated calibration no longer
transfers. A receiver without a model of that accuracy is back to labelled
pairs. Two consequences. The only secret the parties must share
is the key; calibration pairs need never exist as transmitted data, which
closes the pair-supply route of the eavesdropper caveat above (the
below-the-wall access condition remains necessary). And the fully blind
receiver---no key, no labels---was also measured: PCA followed by ICA on the
features of unlabelled messages recovers the components only partially
(mean $|\mathrm{corr}| = 0.47$ one-shot, $0.12$ at the $t=0$ control), and as
a mean-square decoder it stays at chance even after quantile-mapping onto the
known message prior. This is the yield of one attack family, reported as an
existence result about keyless leakage; it is not an upper bound on what a
stronger adversary could extract.

\section{Controls we recommend}\label{sec:checklist}

Each artifact is detectable with a control costing seconds:
\begin{enumerate}
\item \textbf{Never evaluate a readout on the matrix it was fitted on.} If the
  feature dimension exceeds the target dimension, an in-sample residual is an
  interpolation artifact (Artifact~I).
\item \textbf{Use one evaluation routine for all conditions,} and record the
  commit that produced every stored number; a cross-condition claim is only as
  good as the guarantee that both numbers came from the same code path.
\item \textbf{Invert your target analytically before claiming to learn it.}
  If the target is element-wise with public parameters, write down the closed
  form; it is the strongest baseline and it is free (Artifact~II).
\item \textbf{Ablate the quantum features, and check that the encoder does
  work.} If a polynomial in the decoded value matches the full model, or the
  latent code equals a classical map regardless of noise, the reservoir is not
  contributing (Artifacts~II, III).
\item \textbf{Replace the latent variable with an uninformative one}, and
  \textbf{test on held-out messages with the decoder frozen}. If a constant
  code reproduces the result, or held-out performance is at chance, the code
  carries nothing (Artifact~IV).
\item \textbf{Check that your estimate has converged in the number of
  random realizations.} Our own eight-seed median was optimistic by $1.5\times$
  and looked stable; report the estimate as a function of seed count.
\item \textbf{If you compare against a trained model, run its optimizer to
  convergence and report an untrained control of the same family.} Otherwise a
  budget limit reads as a physical conclusion, and an ansatz effect reads as a
  learning effect.
\item \textbf{Report a decoder-free visibility measure} alongside any decode
  error, and \textbf{sweep the input amplitude}. A decode error can improve
  smoothly across a threshold at which some component of the input is in fact
  undetectable [Fig.~\ref{fig:threshold}(a)], while linear visibility can look
  perfect where the decoder is at chance [Fig.~\ref{fig:threshold}(b)].
  Neither quantity alone is safe, and a single amplitude can misstate
  performance by a factor of several.
\end{enumerate}

\section{Discussion}\label{sec:discussion}

The artifacts are independent but mutually reinforcing: the in-sample metric
makes any reservoir look perfect; a classically invertible or empty target
keeps absolute performance from exposing the misattribution; the structurally
noise-free latent code makes the pipeline look robust; and the underdetermined
readout memorizes. Together they produce a coherent story---noise as a
computational resource---that survives internal consistency checks and fails
immediately under a classical control.

What replaces it is narrower but stands. A joint scrambling channel does
transport the input across a reservoir boundary---though a classical
scrambler of matched size does the same task better, so this is a
statement about the geometry surviving the controls, not about a quantum
mechanism. It requires non-Clifford resources beyond mere scrambling, with
the scan of Sec.~\ref{sec:channel} locating where they must sit (in the
scrambler for one-shot injection; supplied by the injection itself when
interleaved), and its feasibility is governed, where a threshold exists at
all, by an access threshold that a decode error does not reveal. Used as a protocol it is keyed
within the transfer experiments run,
it streams arbitrary-length data once the accessible half is dissipatively
refreshed (with a single fixed decoder when the key schedule is stationary),
and its calibration needs no labelled transmissions---the receiver
matches the supervised ceiling to within a few percent from self-simulated
pairs, so the only shared
secret is the key (Sec.~\ref{sec:extensions}). We have not demonstrated an
advantage over classical communication of the same data, and we do not claim
one, nor any cryptographic security beyond the measured access threshold and
key sensitivity; the construction is a proof that the encode--decode geometry
can be made to carry information at all, which is precisely what the withdrawn
version did not establish.

\subsection*{Limitations}

The channel demonstration is at $n = 10$, $\Nc = 4$, in exact statevector
simulation with a single fixed circuit family, and the threshold scans are at
one value of $t$ and one shot budget; the dependence of $f^{*}$ on the doping
convention and on $N$ is untested, and with eight circuit realizations per cell
the individual thresholds carry appreciable scatter---we read the trend with
$\Nc$ and the one-shot/interleaved ordering, not the individual numbers. The response-Gram spectrum is a linear-response
visibility measure and not the quantum Fisher information; more importantly, as
Sec.~\ref{sec:wall} shows, it is a small-signal quantity and therefore does not
by itself predict decodability at the amplitudes a protocol actually uses. We
report it as a necessary condition, not a sufficient one. The finite-shot sweep we ran shows
no magic-driven collapse, but our decoder uses all low-weight Pauli
expectations simultaneously and is therefore the non-adaptive strategy for
which no such collapse is expected: the collapse reported
in~\cite{qnoise} is a per-setting statement rather than a separation in total
measurement resources. Testing it requires a single-setting readout and is
left open. The variational baseline uses one ansatz family, one
gradient-free optimizer and one initialization per seed; a deeper ansatz,
gradient-based training, or restarts could change that comparison, and the
early flattening we observe is consistent with a local optimum. The
hardware run of Sec.~\ref{sec:channel} is a first demonstration at $n=6$
only; its $t=0$ control shows that device-level coherent errors partially
open the Clifford-closed channel, so every doping statement in this paper
is a statement about exact dynamics, not about hardware. The extensions of
Sec.~\ref{sec:extensions} are at $n=8$ with exact expectations except where a
shot budget is stated; the reset is an idealized measure-and-discard rather
than a hardware reset channel; the key-sensitivity matrix samples four keys,
not a keyspace analysis; and the blind-ICA number is the yield of one attack
family, not a bound on an adversary. The digital-twin receiver exists
because at $n=8$ the channel is exactly classically simulable; in a regime
where the reservoir were classically hard to simulate, twin calibration
would not carry over and the label-free result would have to be
re-established by other means. Where key selection is used, the improvement
reflects the large key-to-key variance of random doped-Clifford segments,
not anything streaming-specific; at settings where that variance is small
(the payload-matched static channel at $\Nc=2$) the same selection gains
nothing. The per-observable $1/N$ shot convention favors large access
fractions in principle; measured, the effect is null: recomputing the
nested thresholds under the physical budget (settings growing from $3$ at
$a=1$ to $23$ at $a=8$, state-dependent variances, at both
$N_{\rm tot}=10^4$ and $10^6$) leaves every censored median unchanged,
because near $f^{*}$ the response eigenvalues collapse over orders of
magnitude and a factor-of-$16$ change in the noise does not move the
transition (\texttt{threshold\_physical.json}). Likewise, simulating the
measurement itself---per-setting basis rotation and multinomial sampling,
within-setting covariances included---reproduces the independent-noise
accuracy numbers to within $2$--$10\%$, on the favorable side
(\texttt{sampled\_covariance.json}); the independent-noise convention
used elsewhere is mildly conservative. The extension tables now rest on 8--16 realizations
(streaming and ablation 16, stationary transfer and key matrix 8, twin
mismatch and the stacked-accuracy cells 3--4) against the core channel's
32; our own finding that an eight-seed median
was optimistic by $1.5\times$ (Sec.~\ref{sec:baseline}) still applies to
the smallest of these, and their medians should be read accordingly. The classical controls
show a matched-size classical scrambler decoding the same task better
than the quantum channel; every positive statement in this paper is
therefore about surviving the artifact controls, and none is about
quantum advantage.

\begin{acknowledgments}
We thank the automated adversarial review process that first flagged the
in-sample inconsistency in this manuscript, using models from a different
family than the one that drafted it; the authors had not detected it in prior
internal review. Simulations used Qulacs~\cite{Suzuki2021qulacs} and
Qiskit~\cite{Qiskit2024}.
\end{acknowledgments}

\section*{Data availability}
All code, stored results, and the commands and seeds needed to regenerate
every number are provided in the accompanying repository, including a
self-contained package that reproduces the Artifact~IV controls in about a
minute and the channel and wall scans in a few minutes on one CPU core. The
extension scans of Sec.~\ref{sec:extensions} (streaming and reset ablation,
stationary-key fixed-decoder transfer, key transfer, twin calibration and
blind ICA) are included in the same package with their stored outputs, as
are the adversarial-review repairs (nested-subsystem threshold, magic
$\times$ amplitude scan, classical controls, entanglement probe, twin
mismatch, single-provenance baseline). A claims manifest
(\texttt{results/MANIFEST.md}) maps every number in the positive sections
to the JSON key and script that produced it; the withdrawal-part numbers
(Artifacts I--IV) are reproduced by the separate WQ10255 verification
package rather than this one, and we say so here rather than claim
otherwise.

\bibliographystyle{unsrt}
\bibliography{references}

\end{document}